\documentclass[iop]{emulateapj}
\usepackage{graphicx,epsfig,amssymb,amsmath,color,lineno}

\begin{document}
\bibliographystyle{apj}

\title{OBSERVATIONAL EVIDENCE OF GALAXY ASSEMBLY BIAS}

\author{{\sc Antonio D. Montero-Dorta}\altaffilmark{1}, {\sc Enrique P\'erez}\altaffilmark{1}, {\sc Francisco Prada}\altaffilmark{1}, {\sc Sergio Rodr\'iguez-Torres}\altaffilmark{2,3}, {\sc Ginevra Favole}\altaffilmark{4}, 
{\sc Anatoly Klypin}\altaffilmark{5}, {\sc Roberto Cid Fernandes}\altaffilmark{6}, {\sc Rosa Gonz\'alez-Delgado}\altaffilmark{1},  {\sc Alberto Dom\'inguez}\altaffilmark{7}, {\sc Adam S. Bolton}\altaffilmark{8}, \\
{\sc Rub\'en Garc\'ia-Benito}\altaffilmark{1}, {\sc Eric Jullo}\altaffilmark{9} \& {\sc Anna Niemiec}\altaffilmark{9}}

\slugcomment{Draft; \today}

\shorttitle{Observational evidence of galaxy assembly bias}
\shortauthors{MONTERO-DORTA ET AL.}

\altaffiltext{1}{Instituto de Astrof{\'i}sica de Andaluc{\'i}a (CSIC), Glorieta de la Astronom{\'i}a, E-18080 Granada, Spain; amonterodorta@gmail.com}
\altaffiltext{2}{Instituto de F{\'i}sica Te{\'o}rica, (UAM/CSIC), Universidad Aut{\'o}noma de Madrid, Cantoblanco, E-28049 Madrid, Spain}
\altaffiltext{3}{Campus of International Excellence UAM+CSIC, Cantoblanco, E-28049 Madrid, Spain}
\altaffiltext{4}{European Space Astronomy Center (ESAC), 3825 Villanueva de la Ca~nada, Madrid, Spain}
\altaffiltext{5}{Astronomy Department, New Mexico State University, Las Cruces, NM, USA}
\altaffiltext{6}{Departamento de F\'isica-CFM, Universidade Federal de Santa Catarina, C.P. 476, 88040-900, Florian\'opolis, SC, Brazil}
\altaffiltext{7}{Grupo de Altas Energ\'ias, Universidad Complutense, E-28040 Madrid, Spain}
\altaffiltext{8}{National Optical Astronomy Observatory (NOAO), 950 North Cherry Ave., Tucson, AZ 85719, USA}
\altaffiltext{9}{Laboratoire d'Astrophysique de Marseille - LAM, Universit\'e d'Aix-Marseille \& CNRS, UMR7326, F-13388 Marseille, France}

\begin{abstract}

We analyze the spectra of 300,000 luminous red galaxies (LRGs) with stellar masses 
$M_* \gtrsim 10^{11} M_{\odot}$ from the SDSS-III Baryon Oscillation Spectroscopic Survey (BOSS). 
By studying their star-formation histories, we find two main evolutionary paths converging into 
the same quiescent galaxy population at $z\sim0.55$. Fast-growing 
LRGs assemble $80\%$ of their stellar mass very early on ($z\sim5$), whereas slow-growing LRGs reach 
the same evolutionary state at $z\sim1.5$. Further investigation reveals that their clustering properties on scales of 
$\sim$1--30 Mpc are, at a high level of significance, also different. 
Fast-growing LRGs are found to be more strongly clustered and reside in overall denser large-scale structure 
environments than slow-growing systems, for a given stellar-mass threshold. 
Our results imply a dependence of clustering on stellar-mass assembly history (naturally connected to the 
mass-formation history of the corresponding halos) for a homogeneous population of similar mass and color, which constitutes a strong 
observational evidence of galaxy assembly bias.

\end{abstract}


\section{Introduction}
\label{sec:intro}

\setcounter{footnote}{0}

Luminous red galaxies (LRGs) are broadly considered a homogeneous galaxy 
population, both in terms of color and stellar mass. They are predominantly old 
and quiescent, and their star formation histories (SFHs) resemble that of a passively-evolving 
galaxy population (e.g., \citealt{Eisenstein2003, Maraston2009, Tojeiro2012, Pacifici2016}). 
They are also known to reside at the center of massive dark-matter halos, and are considered 
excellent tracers of the large-scale structure (LSS) of the Universe (e.g., \citealt{Postman1995, Eisenstein2005, White2011, Anderson2014, Rodriguez2016}). 

On the other hand, results from cosmological simulations indicate that the clustering properties of dark-matter halos depend not only 
on halo mass but also on their formation epoch and accretion history
(e.g., \citealt{Gao2005, Wechsler2006, Gao2007, Wang2011, Sunayama2016}). There are reasons to believe that 
this {\it{assembly bias}} manifests itself on the galaxy side as well, so that the clustering 
signal and the properties of galaxies in the LSS are influenced by the accretion history of their host halos (e.g. \citealt{Yang2006, Hearin2013, Zentner2014, Hearin2015, 
Hearin2016, Miyatake2016}). A consensus, however, is yet to emerge, since previous works 
could be affected by differences in halo/stellar mass between galaxy samples and/or contamination by satellite galaxies 
(see, e.g., \citealt{Paranjape2015, Lin2016, Zu2016, Dvornik2017}). 

Here we analyze the SFH and clustering properties of more than 
300,000 LRGs at $0.50<z<0.60$, drawn from the Baryon Oscillation Spectroscopic Survey (BOSS, \citealt{Dawson2013}) of 
the SDSS-III \citep{Eisenstein2011}. Although these galaxies are already quiescent
at these redshifts, we search for evidence of a diverse mass-growth history that could manifest itself in differences
in their clustering signal. The BOSS LRG sample is advantageous in that it maps a galaxy population of similar stellar mass ($M_* \gtrsim 10^{11} M_{\odot}$), which is mostly 
comprised by central galaxies (only $\sim$10$\%$ of satellites) of massive groups and clusters (see, e.g., \citealt{White2011, Rodriguez2016}).

This letter is organized as follows. The data and sample selection are described in Section~\ref{sec:data}. The determination of SFHs for LRGs is addressed 
in Section~\ref{sec:SFH}. Our clustering measurements are presented in Section~\ref{sec:clustering}. Finally, 
in Section~\ref{sec:conclusions}, we discuss the implications of our results and 
summarize the main conclusions of our work. 

Throughout this paper, we adopt a cosmology 
with $\Omega_M=0.307 $,  $\Omega_\Lambda=0.693$ and $H_0 = 100h$ km s$^{-1}$ Mpc$^{-1}$ with $h=0.678$ 
(\citealt{Planck2013}), and use AB magnitudes \citep{OkeGunn1983}.

\section{Data and sample selection}
\label{sec:data}

We use galaxy spectra and photometric data from the Twelfth Data Release of the SDSS 
(DR12, \citealt{Alam2015}), which is the final release of SDSS-III/BOSS. 
We focus on the official data set for cosmological measurements within the 
collaboration, the BOSS DR12 LSS catalog (see \citealt{Alam2015}).
This catalog incorporates a detailed treatment of angular incompleteness and a variety of systematics 
that could potentially affect the target density of spectroscopically-identified galaxies. 
We restrict our analysis to the CMASS (for ``Constant MASS'') sample, containing $\sim$900,000 LRGs 
within the nominal redshift range $0.4<z<0.7$. For a detailed description of the BOSS survey, 
see \cite{Dawson2013}.

In order to maximize stellar-mass completeness and minimize selection effects, we exclude galaxies outside the 
redshift range $0.5<z<0.6$. Below $z\sim0.5$, the red sequence is severely 
incomplete due to the CMASS color-color cuts. Above $z\sim0.6$, the contamination 
from bluer objects in the sample increases significantly (see \citealt{Leauthaud2016, MonteroDorta2016A} for more information on 
 completeness and selection effects). Blue objects within our selected redshift range 
 $0.5<z<0.6$ are further removed by imposing the color cut $g-i > 2.35$ (see \citealt{Masters2011, Maraston2013, Favole2016}).
Our final LRG parent sample comprises a total of 305,741 LRGs, with stellar masses
$M_* > 10^{11}~M_{\odot}$, over an effective area of 9376 deg$^{2}$.

We complement the BOSS data with photometric and morphological information 
extracted from the Data Release 3 (DR3) of the DECam Legacy Survey (DECaLS)\footnote{http://legacysurvey.org/decamls/}. 
DECaLS is an optical survey that will image 6700 deg$^{2}$ to 
a photometric depth of $r=23.9$, i.e., $\sim1.5$ mag deeper than the SDSS imaging. 
The DR3 covers a disjoint footprint of 4200 deg$^{2}$, observed in all three g, r, z SDSS filters. 
DECaLS photometric and morphological information
have been retrieved for $\sim$20$\%$ of our parent sample, i.e., $\sim$55,000 galaxies.

\section{The star formation history of LRG}
\label{sec:SFH}

SFHs and stellar masses for the parent LRG sample 
are determined using the {\sc starlight} code \citep{Cid2005}.
{\sc starlight} fits a spectrum in terms of a non-parametric linear 
combination of a number of single stellar population models (SSPs) from a base spanning different ages and metallicities. An important advantage 
of {\sc starlight} resides in its flexibility in terms of accommodating for any physically-plausible shape for the SFHs.

The base used in this work contains 319 SSPs drawn from the Charlot \& Bruzual CB 2007 library\footnote{http://www.bruzual.org/$\sim$gbruzual/cb07},
where five metallicity values of 0.0004, 0.004, 0.008, 0.02 and 0.05 are considered (Z$_{\odot} = 0.02$). These 
models assume a \cite{Chabrier2003} initial mass function (IMF). Ages range  
from 1 Myr to either 7.5 Gyr (using 63 age bins) or 8.0 Gyr (using 64 age bins), depending 
on the redshift of the galaxy (only ages smaller than the age of the Universe at the corresponding 
redshift are considered). Each spectrum is fitted in the rest-frame wavelength range 3000-5930 \AA.
{\sc starlight} outputs a population vector whose components express the fractional contribution
of each base component to the observed continuum at a reference wavelength of 4450 \AA;
the corresponding mass fractions are also given. Throughout this work, the mass fractions used to compute the stellar 
mass growth are corrected for the mass lost by stars during their evolution. However, the mass fractions used to compute the star formation rates (SFRs)
employ all the mass turned into stars and are therefore not corrected for this evolutionary effect. For the IMF adopted, 
the total mass turned into stars is $\sim$1.93 times the current mass in stars. Finally, dust is accounted 
for using a foreground screen model and a \cite{Cardelli1989} reddening law.

\begin{figure}
\begin{center}
\includegraphics[scale=0.49]{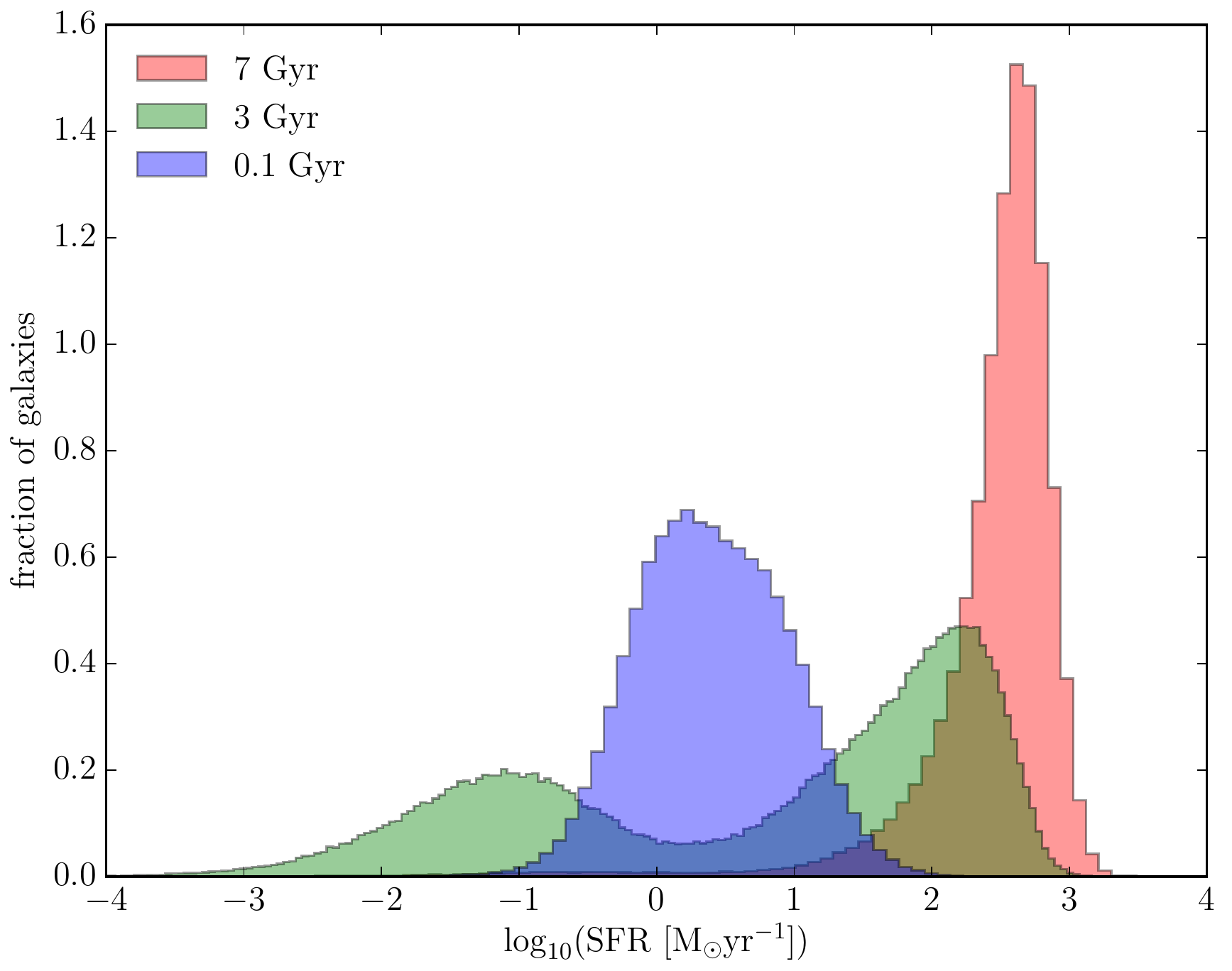}
\caption{The distribution of the logarithm of the SFR, in units of $M_\odot~yr^{-1}$, in three different snapshots of 
galaxy-frame look-back time, centered at 0.1, 3 and 7 Gyr, respectively. The distributions have been normalized
to unit area. The 3-Gyr snapshot is used in this work 
to define two different LRG populations, with measured SFR above and below $2$ $M_\odot~yr^{-1}$, respectively (see text).  } 
\label{fig:selection}
\end{center}
\end{figure}

\begin{figure}
\begin{center}
\includegraphics[scale=0.45]{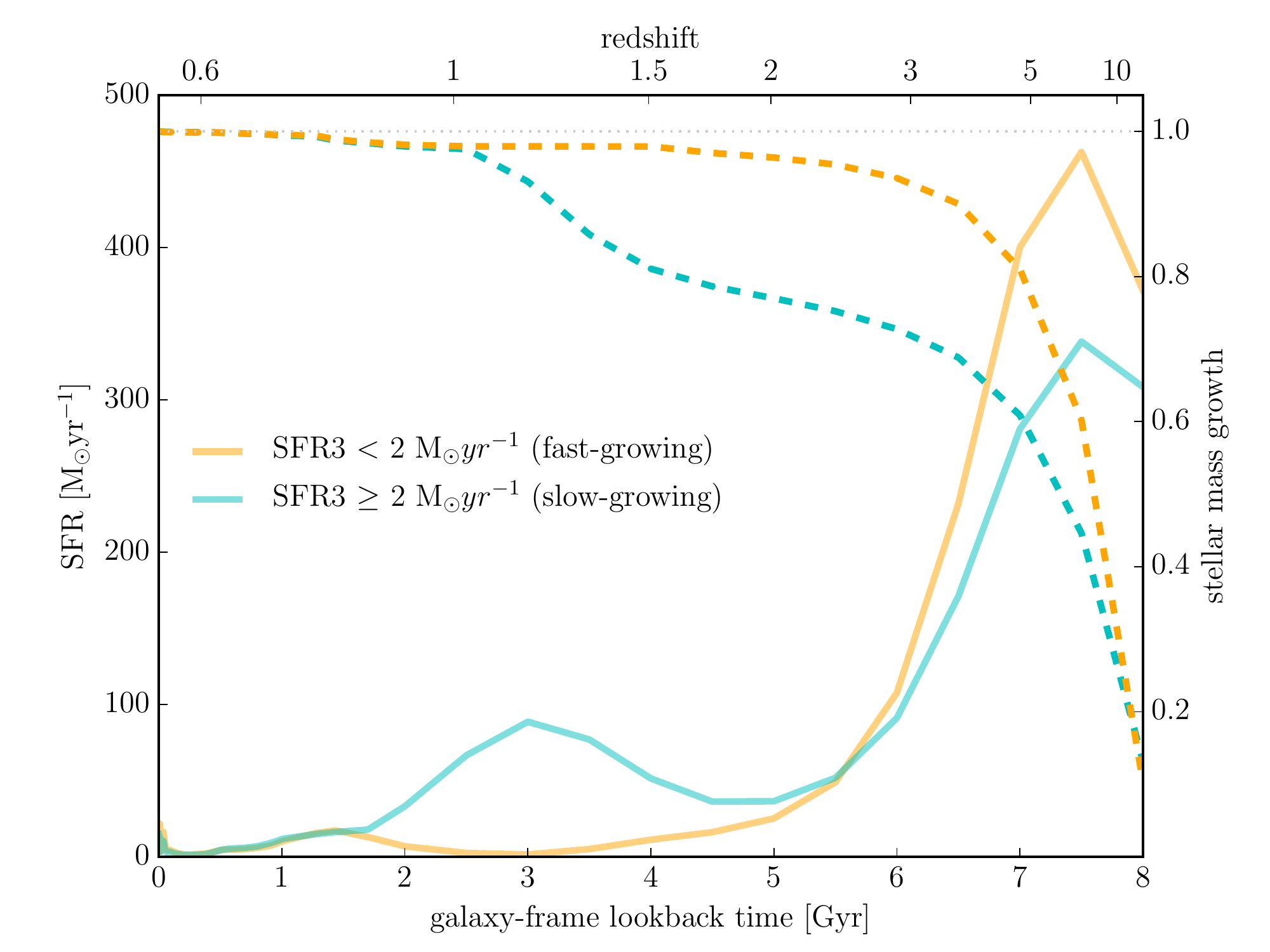}
\caption{Average SFHs (solid) and stellar mass growths (dashed) illustrating the 
two different evolutionary channels for the fast- and slow-growing LRG populations, as defined using the SFR 
at 3 Gyr galaxy-frame look-back time. The corresponding redshift is shown for reference.} 
\label{fig:sfh}
\end{center}
\end{figure}

\begin{figure*}
\begin{center}
\includegraphics[scale=0.65]{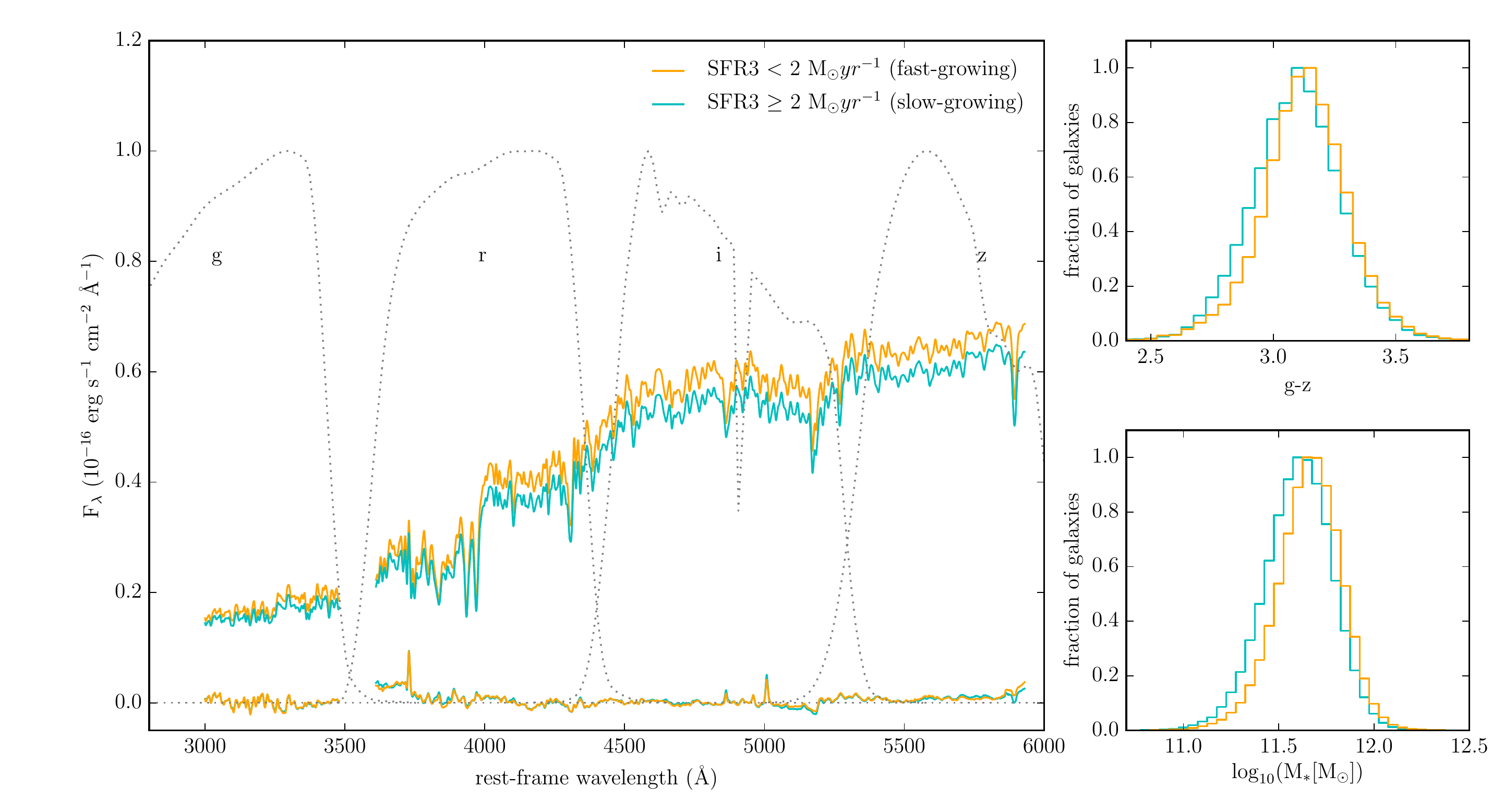}
\caption{{\it Left}: Average rest-frame spectra for the fast- (SFR3 $< 2$ $M_\odot~yr^{-1}$) and the slow-growing (SFR3 $> 2$ $M_\odot~yr^{-1}$)
LRG populations. In the background, the {\it griz} SDSS photometric bands blueshifted by a factor $(1+z)$ are shown for reference. The 
residuals after continuum subtraction are provided at the bottom of the plot. From left to right, the following emission lines are visible:
[OII] doublet, H$\gamma$, H$\beta$, [OIII] doublet. {\it Right (upper)}: $g-z$ color from DECaLS for the two LRG populations. The small shift in the median of both distributions 
confirms the redder shape of the average spectrum for fast-growing LRGs. {\it  Right (lower)}: The distribution of stellar masses for both 
populations. Fast-growing LRGs are $\sim0.06$ dex more massive than slow-growing LRGs.}
\label{fig:spectra}
\end{center}
\end{figure*}

In Figure~\ref{fig:selection}, we show the SFR, in units of $M_\odot~yr^{-1}$, for the parent sample 
in 3 different snapshots of galaxy-frame look-back time\footnote{The age is measured 
retrospectively from $z=0.55$, i.e., 5.5 Gyr ago.} ($t_{back}$). 
These snapshots are centered at 0.1 Gyr ($t_{back} < 0.1$ Gyr), 3 Gyr ($2.5$ Gyr $< t_{back} < 3.5$ Gyr) 
and 7 Gyr ($6.5$ Gyr $< t_{back} < T_{Univ}$; where $T_{Univ}$ is the age of the Universe). Hereafter, the corresponding SFRs
will be named SFR0.1, SFR3 and SFR7, respectively. Figure~\ref{fig:selection} shows a 
narrow distribution for SFR7, the initial star formation (SF) burst, and a bimodal 
distribution for SFR3, with a fraction of LRGs showing signs of mild SF activity while the 
remaining population appears already quiescent. Note that for $\sim20\%$ of the sample, 
the measured SFR3 is strictly equal to 0, since the corresponding ``age components" are not needed to fit the spectra. 
The bimodality observed at 3 Gyr look-back time disappears later on, as the distribution of SFR0.1 indicates. 

We use the bimodal distribution found for SFR3 as a distinctive SFH feature to define two different types of LRGs. 
Galaxies with SFR3 $< 2$ $M_\odot~yr^{-1}$ (49$\%$ of the sample)
are named ``fast-growing" LRGs, whereas objects with SFR3 $\ge 2$ $M_\odot~yr^{-1}$ (51$\%$)
are dubbed ``slow-growing" LRGs. This classification naturally defines two slightly different evolutionary pathways to quiescence, 
as Figure~\ref{fig:sfh} shows. Here, the average SFH along with the average stellar mass growth as a function 
of look-back time is displayed for both populations.  Fast-growing LRGs experience a very 
prominent initial burst, where most of the SF takes place. They form $80\%$ of their mass within 
approximately the first Gyr, i.e., at $z\gtrsim5$. For slow-growing LRGs, the initial burst 
is slightly less powerful, and they experience an episode of SF at $\sim$ 3 Gyr ($z\sim1.5$). Figure~\ref{fig:sfh} shows 
a slower stellar-mass growth for these galaxies: they form $\sim50\%$ of their mass within 
the first Gyr, but it takes them more than 4 Gyr to reach $80\%$ growth. It is
important to bear in mind that these differences in SFH are detectable but small, in light of the known
uncertainties in stellar population modeling. 

It is noteworthy that the existence of multiple paths to quiescence for massive red galaxies has been extensively discussed
in the literature (see, e.g., \citealt{Fritz2014, Pacifici2016, Henriques2016}). Evidence of recent SF activity, indicating 
small deviations from purely-passive evolution similar to those reported here, are well documented (e.g., \citealt{Tojeiro2012,Fritz2014, Citro2016}).

\begin{figure*}
\begin{center}
\includegraphics[scale=0.64]{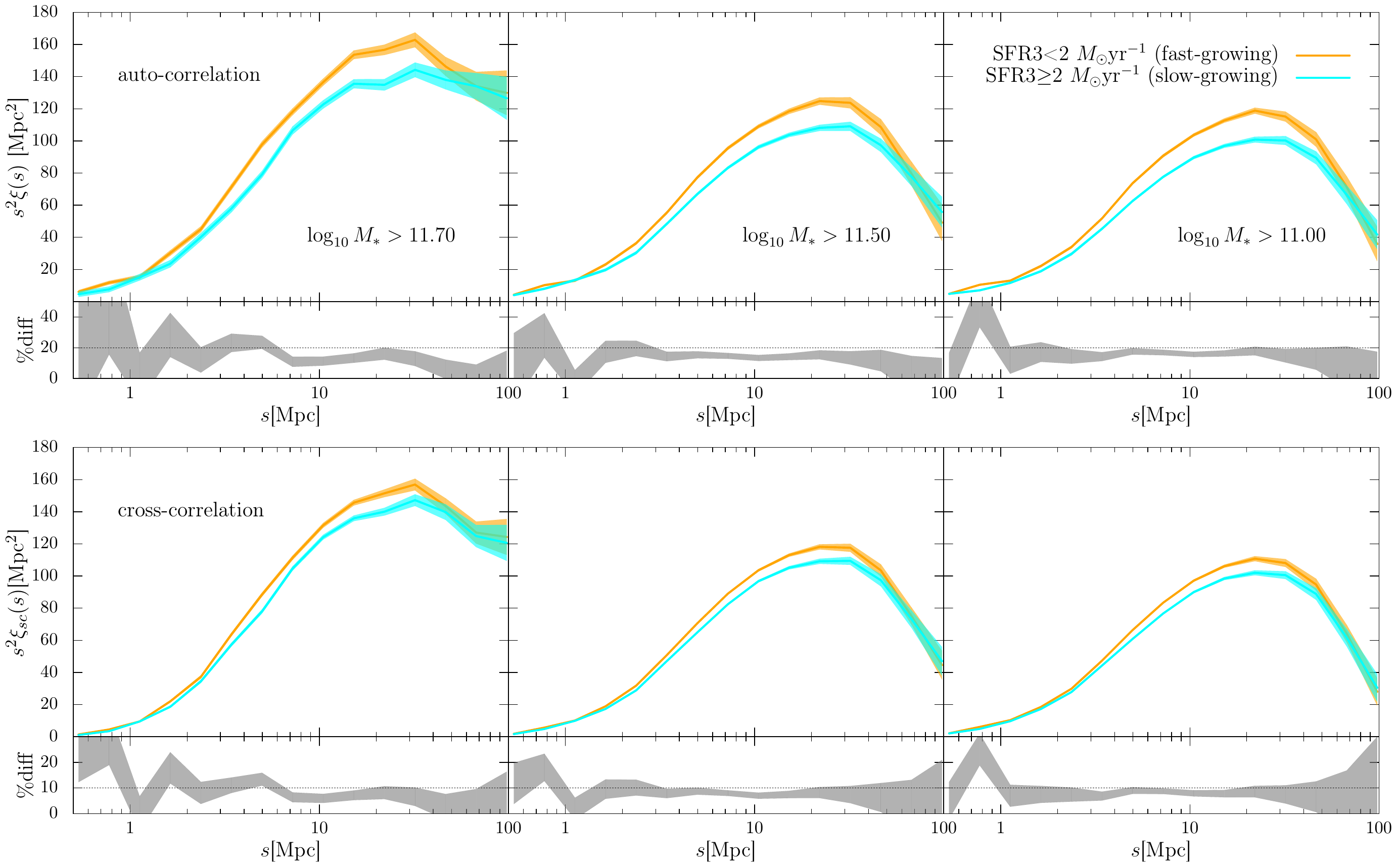}
\caption{Clustering properties of LRGs for different stellar-mass thresholds. {\it Top}: From left to right, the monopole of the redshift-space 2D correlation function (auto-correlation) for the fast- and 
the slow-growing LRG populations in cumulative stellar mass bins of $\log_{10} M_* (M_\odot)> $ 11.7, 11.5, and 11, 
respectively. {\it Bottom}: The cross-correlation between each of the LRG populations and the entire parent sample, for the same cumulative stellar-mass bins.
In both panels, the relative difference between the two functions is shown in the subplots. Error bars are computed using a set of BOSS DR12 MultiDark-Patchy mocks.
Fast-growing LRGs are $\sim20\%$ more clustered and reside in overall denser environments on all scales below $\sim 30$ Mpc.} 
\label{fig:2PCF}
\end{center}
\end{figure*}

Fast- and slow-growing LRGs present small but noticeable 
differences in several other properties. This is illustrated in 
Figure~\ref{fig:spectra}. In the left-hand panel, the average spectra 
for both populations is presented in rest-frame. Fast-growing LRGs are slightly redder than 
their slow-growing counterparts. This difference is also noticeable and consistent with the $g-z$ color distribution 
displayed in the top-right panel of Figure~\ref{fig:spectra}. Here, we use DECaLS photometry for 
our crossmatched sample of $\sim$55,000 objects. The difference in the median $g-z$ color between fast- and slow-growing 
LRGs is $0.041$ mag. Both populations contain emission-line objects, as the residuals after continuum subtraction in the left-hand 
panel of Figure~\ref{fig:spectra} demonstrate. However, no significant difference in emission-line properties has been detected between samples.

Importantly, our LRG classification has little impact on stellar mass, as shown in the bottom-right
panel of Figure~\ref{fig:spectra}. Fast-growing LRGs are on average slightly more massive 
than their slow-growing counterparts, but only by $0.058$ dex (as measured 
from the median values). This difference is small considering the uncertainties in the determination 
of stellar masses. We have checked that 
the stellar masses computed using {\sc starlight} for the BOSS CMASS sample are consistent with previous estimates from 
the Granada FSPS (\citealt{Ahn2013}), Portsmouth (\citealt{Maraston2013}) and Wisconsin PCA (\citealt{Chen2012}) galaxy products.

As expected, slow-growing LRGs present also younger stellar populations at $z=0.55$; the flux-weighted mean age is 2.92 Gyr, as compared 
to 3.30 Gyr for fast-growing systems. A detailed study on the stellar population properties of CMASS LRGs obtained using the 
{\sc starlight} code is currently in preparation. In addition, 
a morphological analysis of LRGs using DECaLS will be presented in Favole et al. (in prep.). In this regard, no significant 
differences in terms of morphology have been found between fast- and slow-growing LRGs. We anticipate that 
$83\%$ of the sample is well described by a De Vaucouleurs light profile. The remaining fraction 
follows either an exponential or a composite profile.

\section{Clustering analysis}
\label{sec:clustering}

The clustering properties of the two LRG populations discussed in Section~\ref{sec:SFH} 
have been analyzed using the two-point correlation function (2PCF). The 2PCF is defined as
the excess probability, compared with that expected for a random distribution, of finding a pair of galaxies at a
given separation. We focus here on the monopole of the 2D correlation function in redshift-space, 
$\xi(r_p,\pi)$, where $s=\surd{r_p^2+\pi^2}$ ($r_p$ is the perpendicular component to the line-of-sight and $\pi$ is the parallel component). 
We use the \cite{Landy1993} estimator to compute this function. Random 
catalogues 20 times larger than our data samples are employed.
For a detailed description of this procedure, see \cite{Rodriguez2016}.

\begin{figure}
\begin{center}
\includegraphics[scale=0.61]{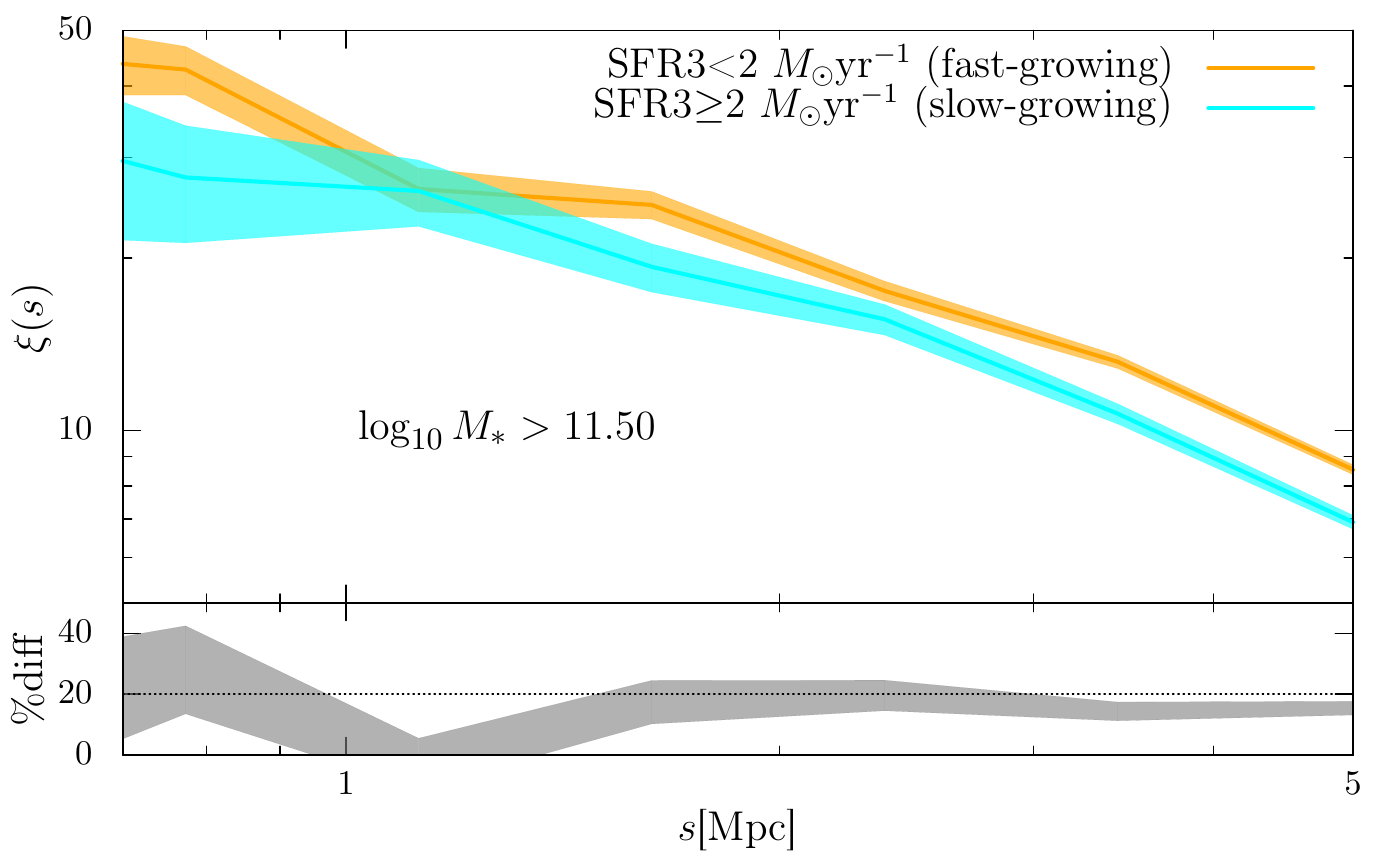}
\caption{The monopole of the redshift-space 2D correlation function (auto-correlation) on small scales for the fast- and 
the slow-growing LRG populations in the cumulative stellar mass bin $\log_{10} M_* (M_\odot)> $ 11.5. Error bars 
are computed using a set of BOSS DR12 MultiDark-Patchy mocks. Fast-growing LRGs have
  $\sim 20\%$ stronger clustering amplitude on scales $s\gtrsim1$ Mpc. Increasing level of noise makes it difficult to measure the
  correlation function reliably on even smaller scales, although the data are
  consistent with the same trend even below 1 Mpc.} 
\label{fig:small_scales}
\end{center}
\end{figure}

The top panel of Figure~\ref{fig:2PCF} displays the monopole of the redshift-space correlation function for the fast- and 
slow-growing LRG populations in cumulative stellar-mass bins of $\log_{10} M_* (M_\odot)> $ 11, 11.5, and 11.7. 
Completeness in stellar mass is greater than $80\%$ for the latter, as measured from the 
stellar mass function of the CMASS LRG sample (see \citealt{Rodriguez2016}). Errors on these estimates 
are computed using a set of BOSS DR12 MultiDark-Patchy mocks \citep{Kitaura2016}. Figure~\ref{fig:2PCF} 
shows that the amplitude of the monopole for fast-growing LRGs is $\sim20\%$ larger than 
that of the slow-growing population, on scales between $\sim$1 and 30 Mpc, independently 
of the stellar-mass threshold adopted. This result is statistically significant at a $\sim$5-$\sigma$ level at $\sim$ 15 Mpc, according to our 
error estimates. A zoom-in on the correlation function at small scales ($s \le 5$ Mpc) for the intermediate mass 
bin is provided in Figure~\ref{fig:small_scales}. 
This figure demonstrates that the amplitude of clustering is 
systematically larger for fast-growing LRGs down to scales of $\sim$1 Mpc, or even below.

A dependence of the clustering signal on stellar mass is noticeable in the top panel of Figure~\ref{fig:2PCF}, as expected, since
more massive LRGs are hosted by larger dark-matter halos (see \citealt{Rodriguez2016}). Yet,
when the sample is split in fast- and slow-growing LRGs for a given stellar-mass threshold 
(equivalent to a given halo-mass threshold, see \citealt{Behroozi2013}), we observe a clear dependence 
of the spatial distribution of LRGs (and hence of their dark-matter halos) on their SFHs. 

As mentioned in Section~\ref{sec:SFH}, fast- and slow-growing LRGs have very similar stellar-mass distributions, with a 
median difference of only $\sim0.06$ dex. In order to quantify the effect of these stellar-mass differences 
on the clustering signal shown in Figure~\ref{fig:2PCF}, we have carried out two separate tests (100 realizations each). In the first test, 
we impose the stellar-mass distributions of both LRG populations to be exactly the same, by randomly removing
galaxies from each subsample. In the second test, we randomly generate pairs of subsamples having 
the same stellar-mass distribution as each of the LRG populations, but now independently of their SFHs. 
Results from the first test show that the difference in the clustering amplitude
decreases slightly, but remains significant within the uncertainties ($\sim15\%$), 
after the stellar-mass dependence has been removed. This suggests that an additional parameter,
related to the SFH (or to the stellar-mass assembly history), is 
necessary to explain the differences seen in Figure~\ref{fig:2PCF}. The second test confirms 
this hypothesis, since the amplitude for the two sets of randomly-generated LRG
populations differ in less that $5\%$ when the dependence on SFH is removed. These tests 
rule out the possibility that stellar mass is responsible for the difference seen in the clustering properties of fast- and
slow-growing LRGs, which constitutes a clear manifestation of galaxy assembly bias.

In short, we find that fast-growing LRGs are more tightly clustered than their slow-growing 
counterparts, and that this effect is not due to the small stellar-mass differences found between the two populations. 
In order to further determine whether fast-growing LRGs reside in denser LSS environments, we compute and compare
the cross-correlation between each LRG population and the entire sample. Results are displayed in the bottom panel of Figure~\ref{fig:small_scales}, 
for the same stellar-mass bins discussed above. The amplitude of the cross-correlation function is $\sim10\%$ larger at scales between $\sim$1 and 30 Mpc
for fast-growing LRGs, in all stellar-mass bins, which confirms that these galaxies live in 
overall denser environments than slow-growing systems.

\section{Discussion and conclusions}
\label{sec:conclusions}

The stellar-population analysis of LRGs presented in this work shows 
two main evolutionary channels converging into the same quenched population at $z=0.55$: 
fast-growing LRGs assemble the majority of their
stellar mass very early on, while the remaining population experience a slower growth. 
Although the differences in SFH are relatively small, the two populations have 
significantly different clustering amplitudes ($\sim20\%$) at scales between $\sim$1 and 30 Mpc, in the sense that fast-growing LRGs are 
found to be more strongly clustered than their slow-growing counterparts. 
Fast-growing LRGs are also found to reside in overall denser LSS environments. 

The observed difference in the clustering amplitude, at two-halo term scales of 
$\sim$5-30 Mpc, reveals a halo-bias ratio of $\sim10\%$, which is of the order 
of the expected bias dependence on halo concentration (or halo formation time), for a 
given halo mass (see \citealt{Wechsler2006, Hearin2013, Hearin2016}). 

We have checked that the differences in the clustering amplitude
cannot be explained by small differences in stellar mass between the two populations.
Our results thus support the hypothesis of galaxy assembly bias, which states that 
the clustering and properties of galaxies depend, not only on the mass of their
host halos, but also on their accretion history. In terms of halo-galaxy modeling, age-distribution-matching techniques have 
succeeded in reproducing the clustering of red and blue galaxies separately, at fixed halo 
mass \citep{Hearin2013, Hearin2014}. Similar results have been obtained using  
``decorated" halo occupation models \citep{Hearin2016}. Yet, no previous study has focused 
on a homogeneous galaxy population of similar stellar mass, such as LRGs. Our results are unique in that they 
present the first direct link between clustering and the stellar-mass assembly history of massive galaxies. 
A detailed modeling of our clustering results including age-matching in the halo abundance 
matching prescription and weak-lensing constraints will be presented in a forthcoming paper.  

Our results are consistent with previous works that show that massive red galaxies at $z\lesssim1$ become
quiescent more rapidly in denser environments, which implies that the overall quenching 
efficiency depends on the development of large-scale structure
(see, e.g., \citealt{Peng2010,Darvish2016,Faisst2017,Henriques2016} for discussion). 

The observational evidence of galaxy assembly bias reported in this work has fundamental implications 
for the modeling and interpretation of LSS galaxy survey data that use galaxy clustering to 
extract cosmological information from the underlying matter-density field.

\section*{Acknowledgments}

ADMD, FP, SRT, GF and AD acknowledge support from the Spanish MICINNs Consolider-Ingenio 2010 Programme 
under grant MultiDark CSD2009-00064 and MINECO grant AYA2014-60641-C2-1-P. RGD, EP, and RGB are supported by MINECO 
grants AYA2016-77846-P and AYA2014-57490-P and Junta de Andalucía P12-FQM-2828. We acknowledge the support of the 
IAA Computing Center staff for the use of the IAA-CSIC computing grid, where all the {\sc starlight} fits have been performed.
A.D. thanks the support of the Juan de la Cierva program from the Spanish MEC. 

Funding for SDSS-III has been provided by the Alfred P. Sloan Foundation, the Participating Institutions, the National Science 
Foundation, and the U.S. Department of Energy Office of Science. The SDSS-III web site is http://www.sdss3.org/.


\end{document}